\documentclass[twocolumn,showpacs,amsmath,amssymb,floatfix]{revtex4}

\usepackage{nicefrac} \usepackage{times} \usepackage{graphicx}
\usepackage{dcolumn} \usepackage{bm}

\DeclareMathOperator{\sign}{sgn}
\begin{document}

\preprint{APS/123-QED}

\title{Movers and shakers: Granular damping in microgravity}

\author{M.~N.~Bannerman} 
\author{J.~E.~Kollmer} 
\author{A.~Sack}
\author{M.~Heckel} 
\author{P.~Mueller} 
\author{T.~P\"oschel}
\affiliation{ Institute for Multiscale Simulation, Universit\"at
  Erlangen-N\"urnberg, N\"agelsbachstra\ss{}e 49b, 91052 Erlangen, Germany }

\date{\today}

\begin{abstract}
  The response of an oscillating granular damper to an initial
  perturbation is studied using experiments performed in microgravity
  and granular dynamics simulations. High-speed video and image
  processing techniques are used to extract experimental data. An
  inelastic hard sphere model is developed to perform simulations and
  the results are in excellent agreement with the experiments. The
  granular damper behaves like a frictional damper and a linear decay
  of the amplitude is observed. This is true even for the simulation
  model, where friction forces are absent. A simple expression is
  developed which predicts the optimal damping conditions for a given
  amplitude and is independent of the oscillation frequency and
  particle inelasticities.
\end{abstract}

\pacs{Valid PACS appear here}

% \keywords{Suggested keywords}

\maketitle
%%%%%%%%%%%%%%%%%%%%%%%%%%%%%%%%%%%%%%%%%%%%%%%%%%%%%%%%%%%%%%%%%%%%
%%%%%%%%%%%%%%%%%%%%%%%%%%%%%%%%%%%%%%%%%%%%%%%%%%%%%%%%%%%%%%%%%%%%
\section{Introduction}
The characteristic property of dynamic granular systems, when compared
to other many-particle systems, is their ability to dissipate
mechanical energy through particle collisions. While the dissipative
properties of vibrated granulate have long been
investigated~\cite{Clara1,Clara2}, recently a large body of
literature~\cite{MaoEtAl:2004a,MaoEtAl:2003,MaoWangXuEtAl:2004,TianningEtAl:2001,BaiKeerWangSnurr:2009,BaiShahKerrEtAl:2008}
has emerged on the mechanics and technical application of this damping
mechanism in the form of {\em granular dampers}.  A granular damper is
a container partly filled by granular particles which may be attached
to vibrating machinery to attenuate the amplitude of the
oscillations. In its regime of operation, the granular material is in
a gaseous state and its dynamics is determined primarily by the
interparticle collisions rather than by long-lasting sliding contacts
between the grains. {\em Static granular dampers} (e.g.,
Refs.~\cite{BourinetHoueedec:1999,CowdenEtAl:2003}) which exploit the
rheology of granular matter and {\em impact dampers} (e.g.,
Refs.~\cite{Cempel:1975,Masri1970,Masri:1969,DuncanWassgrenKrousgrill:2005}),
where only one or few particles are located in a cavity and dissipate
energy in collisions with the walls of the container, are not
considered here.

Granular dampers have a number of properties which are desirable in a
wide range of technical applications: Unlike traditional dampers,
granular dampers do not require an anchor in order to restrict the
motion of the system. This is advantageous for damping in portable
equipment and in space applications where no fixed anchor is
available. Granular dampers are extremely simple devices consisting
solely of particles enclosed in a container or cavity and require very
little maintenance. Granular dampers do not suffer from significant
aging when compared to the oil and rubber components of traditional
dampers. Finally, granular dampers can operate over a wide range of
temperatures without performance degradation as the mechanics of the
particle-particle and particle-wall interactions exhibit only a weak
dependence on the temperature.  Modern technical applications of
granular dampers include the damping of blade integrated disks
(blisks) for compressors~\cite{KielbEtAl:1998}, structural vibration
damping~\cite{Panossian:1992,CempelLotz:1993,Fricke:2000}, noise
reduction of bank note processing machines~\cite{Xu:2004a} and
others. Perhaps the most common application is the {\em dead-blow
  hammer}~\cite{deadblowhammer:1967} and in other impact damping
handles~\cite{Ashley:1995}.% \cite{Wijk_1982}.

The macroscopic damping properties of granular dampers under dynamic
load is complicated, highly non-linear, and there is no
straightforward way to optimize their performance for a given
situation. This has been demonstrated in a number of experiments and
Molecular Dynamics (MD) simulations, including investigations on the
attenuation of a free spring or cantilever with an attached granular
damper~\cite{MaoWangXuEtAl:2004,BaiKeerWangSnurr:2009,KinraMarhadiWitt:2005,Marhadi:2003,FriendKinra:2000}.
The response of an oscillating cantilever with respect to periodic
forcing has also been
studied~\cite{Saeki:2002,Saeki:2005,FowlerFlintOlson:2000,FowlerFlintOlsen:2001,PapalouMasri:1998,Wong_etal_2009}.
Even more complex systems have been investigated, such as the
oscillation modes of a plate with an abundance of granulate filled
cavities~\cite{XuWangChen:2004,Xu:2005,ParkPalumbo:2009,Fowler:2003}
with the aim of noise reduction~\cite{Xu:2004a}.  For simple systems,
such as cantilever oscillators, some progress has been
made. Theoretical models have been developed based on phenomenological
descriptions of the multiphase gas-particle flow of granular matter
for attenuating oscillations~\cite{WuLiaoWang:2004} and also for
driven steady state oscillations~\cite{Fang:2006}.

A granular damper, that is a dynamical system of dissipative
interacting particles, obviously must be able to dissipate energy;
however, its general behavior is not clear {\em \'a
  priori}. Properties, such as the dissipation rate, are complex
functions of the frequency and amplitude of the oscillation, as well
as the particle properties, the extension and characteristics of the
container or cavities, and the filling fractions. More work is needed
in this field to generate experimental results and corresponding
models capable of describing the dynamics of granular dampers.

Saluena et al.~\cite{Clara1} have shown that several regimes of energy
dissipation exist for a granular damper and that the transitions
between these regimes are determined primarily by the influence of
gravity. An efficient operation of a granular damper can be expected
only if the average kinetic energy of the particles is much larger
than their average potential energy and the damper operates in the
{\em dynamic or collisional regime}. In order to carefully investigate
this regime, the influence of gravity should be minimized and
experimental investigations should be performed under conditions of
weightlessness.

The objective of this paper is to develop an effective model for the
energy dissipation of a granular damper operating in the collisional
regime. Our approach is as follows: First, experiments in microgravity
are performed and the attenuation of a spring with an attached
granular damper for several sets of parameters is obtained
(Sec.~\ref{sec:experiment}). A model capable of reproducing the
experimental results is also developed and high-precision Discrete
Element Method (DEM) simulations are performed
(Sec.~\ref{sec:simandmodel}). The two free parameters of the model
(inelasticities) are obtained by adjusting the values until the
simulation matches the experiment as closely as possible for a single
experiment (Sec.~\ref{sec:parameters}). From the excellent agreement
of the simulation results for the fitted system and {\em for all
  other} experiments, it is concluded that the model underlying the
simulation replicates the system's essential features
(Sec.~\ref{sec:validation}). Thus, the DEM simulations are an
effective model for granular damping in the collisional regime.  In
Sec.~\ref{sec:optimisation}, a simple equation for the optimal design
of a simple damper is derived and tested against the results of the
DEM simulations. Section~\ref{sec:macmodel} discusses the observed
linear decay of the amplitude and compares it to friction-damped
oscillators. Finally, in Sec.~\ref{sec:conclusions} the conclusions of
the paper are outlined.

%%%%%%%%%%%%%%%%%%%%%%%%%%%%%%%%%%%%%%%%%%%%%%%%%%%%%%%%%%%%%%%%%%%%
%%%%%%%%%%%%%%%%%%%%%%%%%%%%%%%%%%%%%%%%%%%%%%%%%%%%%%%%%%%%%%%%%%%%
\section{Experimental Setup\label{sec:experiment}}

\begin{figure}[ht]
  \centerline{\includegraphics[clip,width=\columnwidth]{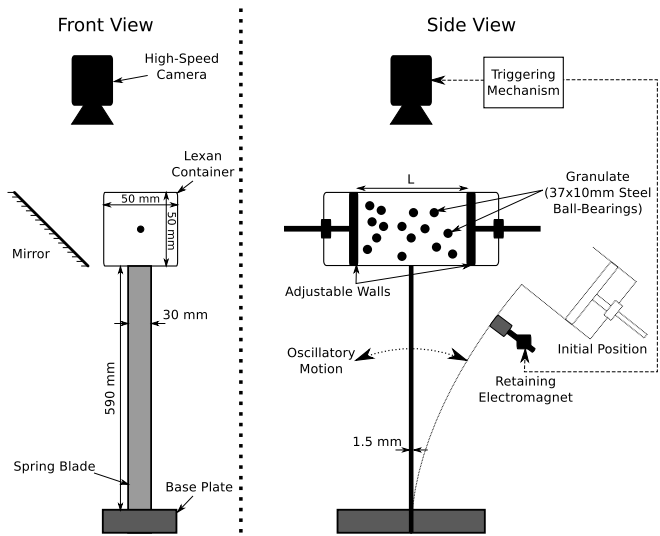}}
\caption{\label{fig:experimentaldesign} 
  Schematic of the experimental setup in front view (left) and side
  view (right). The curvature of the oscillations is exaggerated for
  the purpose of illustration. }
\end{figure}

Figure~\ref{fig:experimentaldesign} is a diagram of the experimental
setup. Our granular damper comprises of a container of adjustable
length which is partially filled with granular material. The damper is
mounted to one end of a spring-steel blade and the opposite end is
clamped in a solid aluminum base plate. The spring blade is described
fully in Sec.~\ref{sec:parameters}. The rectangular damper container
is constructed from 5~mm thick transparent polycarbonate plates. The
internal dimensions of the container are 50~mm$\times$50~mm$\times L$,
where the length $L$ (in the direction of the oscillation) is adjusted
by altering the spacing of the end walls. The container's net weight
(without granulate) is $M=434$~g. In this work, four different
container lengths of $L=40,65,85,$ and $104$~mm are used. The damper
is loaded with 37 precision steel ball-bearings of diameter
$\sigma=10$~mm and mass $m=4.04$~g. This number of particles is chosen
as it packs to form a dense layer, two particles deep, on the end
walls of the container.

The motion of the damper and contained granulate is recorded using a
high speed camera (MotionScope M3\texttrademark), which records at a
frame rate of 500~fps and with a spatial resolution of
$1024\times1280$ pixels. A $45^\circ$ mirror is placed at the side of
the container and allows for the simultaneous observation of the
granulate from the top and the side. The position of the damper and
the center of gravity of the particles are extracted from the top-view
using standard image processing techniques. Although the side view
facilitates more complex methods of reconstruction, it will be shown
that the motion of the granulate is well described by the center of
mass motion. All position measurements are made in a 2D plane which is
parallel to and intersecting the top of the container while it is in
its equilibrium position. Although this introduces some error at the
peaks of the oscillations due to the curved damper trajectory, this
error is negligible.

At the start of the experiment, the spring blade is deflected and held
at the initial displacement of $\Delta_0=107.5$~mm using an
electromagnet.  A trigger mechanism begins the experiment and starts
the camera recording. After a short delay of 1~s, the spring blade is
released from the electromagnet and the oscillations are recorded for
30 seconds.

To assure conditions of weightlessness, the experiment is performed on
a modified Airbus A300 aircraft which has been retrofitted for
performing parabolic flights.  The parabolic flight provides a
suitable microgravity environment ($\pm0.05$~g) which lasts around
22~s and allows a number of experiments to be performed. In the
following section, the numerical model and simulation techniques are
described.
%%%%%%%%%%%%%%%%%%%%%%%%%%%%%%%%%%%%%%%%%%%%%%%%%%%%%%%%%%%%%%%%%%%% 
%%%%%%%%%%%%%%%%%%%%%%%%%%%%%%%%%%%%%%%%%%%%%%%%%%%%%%%%%%%%%%%%%%%%
\section{Numerical Model and Simulation Method}
\label{sec:simandmodel}
A sufficiently complex model, capable of reproducing the observed
experimental behavior, must be found if the system's dynamics is to be
understood. The model presented here is complex enough to yield
quantitative agreement with the experiments and yet simple enough to
gain insight into the dynamics of the system. The model for the
granulate consists of a system of $N$ smooth inelastic hard spheres,
each of mass $m$ and diameter $\sigma$. Although inelastic hard
spheres are a basic model for the granulate they capture all of the
essential behavior of the system: dissipative interactions between
hard spherical particles. Friction forces, which typically play a
large role in granular systems, are also neglected and the
implications of this approximation are discussed later.

To model the oscillating mass and granular damper, the hard spheres
are shaken in a rectangular box of mass $M$, which is coupled to a
linear spring. The box is assumed to remain parallel to the axis of
the system and oscillate along only one axis. By only permitting
oscillations in a single dimension, this model neglects the arcing
motion of the blade spring (see Fig.~\ref{fig:experimentaldesign}) and
oscillatory modes induced by the collisions of the particles with the
box; however, these effects are expected to be small in comparison to
the dynamics of the modeled oscillation.  With these approximations,
the collision-free motion of the box can then be modeled using a
simple harmonic oscillator
\begin{align}
  \vec{r}_{\rm Box}(t) = \hat{n}\, \Delta\, \cos\left[2\, \pi\,
    \omega\left(t + t_{\rm shift}\right)\right] + \vec{r}^{\,(0)}_{\rm
    box}
\end{align}
where $\vec{r}_{\rm Box}$ is the current position of the oscillator,
$\vec{r}_{\rm Box}^{\,(0)}$ is its equilibrium position, $\hat{n}$ is
the unit vector in the direction of the oscillation, and $\omega$ is
the frequency of the empty damper. The amplitude of the oscillation
$\Delta$ and the phase shift of the oscillator $t_{\rm{}shift}$ are
dynamical quantities and are altered by particle-box interactions.  At
time $t=0$, the plate is at its positive maximum ($t_{\rm{}shift}=0$)
with an initial displacement of $\Delta=\Delta_0$.

The methods for performing event-driven simulations using smooth
hard-spheres and fixed walls are well established (e.g., see
Ref.~\cite{Poschel_Schwager_2005}) and will not be discussed in
detail. Here an event-driven dynamics simulation package ({\sc
  Dynamo}~\cite{DYNAMO}) is used to simulate the dynamics of
granular-damped oscillators.  The only extension to the basic
event-driven method concerns the detection and execution of events
between particles and the oscillating boundary walls perpendicular to
the oscillation direction $\hat{n}$, which is discussed in the
following subsections.
%%%%%%%%%%%%%%%%%%%%%%%%%%%%%%%%%%%%%%%%%%%%%%%%%%%%%%%%%%%%%%%%%%%%
\subsection{Detecting Oscillating Wall Interactions}
Event-driven algorithms require an expression to calculate if and when
a collision (an {\em event}) occurs between a particle and the
bounding walls of the damper. If a collision is detected and it is the
next event to occur in the system, the system is moved to the time of
the collision and the event is executed by updating the velocities of
the colliding particle, and the phase shift $t_{\rm shift}$ and
amplitude $\Delta$ of the oscillator.

To determine the time at which a particle $i$ will collide with an
oscillating wall, the equations of motion for the particle and the
oscillating plate must be solved. Essentially, this is a search for
the shortest positive root $\Delta t$ of the function
\begin{align}\label{eq:f0}
  f(\Delta t) = \left[\vec{r}_i + \Delta t\,\vec{v}_i - \vec{r}_{\rm
      Box}(\Delta t+t)\right] \cdot \hat{n}\pm
  \left(\frac{L-\sigma}{2}\right) = 0
\end{align}
where $\vec{r}_i$ and $\vec{v}_i$ are the position and velocity of
particle $i$ at the current system time $t$, and $\Delta t$ is the
time till collision.  The sign of the term $\pm (L-\sigma)/2$ is used
to set which side of the oscillating box is tested for collisions.

To guarantee that no roots are missed, the root finding technique of
Frenkel and Maguire~\cite{Frenkel_Maguire_1983} is used.  This root
finding routine requires a fixed interval to search for possible
roots. The upper bound on the interval to search is determined from
the time the freely moving particle takes to cross the extrema of the
tested-wall's oscillation
\begin{align}
  \Delta t_{\max} = \frac{\sign(\hat{n} \cdot \vec{v}_i)
    \left(\left[L-\sigma\right]/2 + \Delta\right) - \hat{n}
    \cdot \left(\vec{r}_i-\vec{r}^{\,(0)}_{\rm
        Box}\right)}{\hat{n} \cdot \vec{v}_i}
\end{align}
where $\sign(x)$ is the sign function.  The lower bound is typically
the current system time ($\Delta t_{\min} = 0$); however, if the last
event to occur was a collision between this particle and a oscillating
wall, the lower bound is increased to avoid re-detecting the same
root. The lower bound is then set to
\begin{align}
  \label{eq:1}
  \Delta t_{\min}=\frac{\left|2\,\dot{f}(0)\right|}{\ddot{f}_{\max}}
\end{align}
where $\ddot{f}_{\max}=\Delta\,\omega^2$ is the maximum absolute
second derivative of Eq.~\eqref{eq:f0}.  The root finding technique
used to search for suitable roots of Eq.~\eqref{eq:f0} iterates
towards a root from the boundaries of the interval by approximating
the function at each iteration with a parabola. The equation of the
parabola is generated using the derivatives of Eq.~\eqref{eq:f0} and
its smallest root provides the next iteration point. The iterations
are halted on the $n$th iteration once the following criterion is met
\begin{align}
  \left|\Delta t_n - \Delta t_{n-1}\right| <
  \frac{\left[L-\sigma\right]}{2\,\dot{f}_{\max}}\times10^{-12}
\end{align}
where
$\dot{f}_{\max}=\left|\vec{v}_i\cdot\hat{n}\right|+2\,\pi\,\omega\,\Delta$
is the maximum absolute first derivative of Eq.~\eqref{eq:f0}. Unlike
the hard line system of Frenkel and Maguire
\cite{Frenkel_Maguire_1983}, all roots of Eq.~\eqref{eq:f0} are
acceptable and only the earliest root must be found. This completes
the description of the collision detection and root finding technique.
%%%%%%%%%%%%%%%%%%%%%%%%%%%%%%%%%%%%%%%%%%%%%%%%%%%%%%%%%%%%%%%%%%%%
\subsection{Executing Particle-Oscillating Wall Collisions}
The final part of the simulation algorithm concerns the execution of
oscillating boundary wall collisions. The conservation of momentum and
the assumption of a constant coefficient of inelasticity leads to
\begin{align}
  \Delta \vec{p}_i = -\Delta \vec{p}_{\rm wall} = - \mu
  \left(1+\varepsilon_{\rm pw}\right)\left(\hat{n} \cdot
    \left[\vec{v}_i - \dot{\vec{r}}_{\rm
        Box}\right]\right)\,\hat{n}
\end{align}
where $\Delta\vec{p}_i$ and $\Delta\vec{p}_{\rm wall}$ are the
momentum change of the colliding particle $i$ and oscillating wall
respectively, $\mu=m\,M/(m+M)$ is the relative mass, and
$\varepsilon_{\rm pw}$ is the coefficient of inelasticity for
particle-wall collisions. During a collision, the phase,
$t_{\rm{}shift}$, and amplitude, $\Delta$, of the oscillating wall are
altered under the constraints of conserving momentum and the current
box position. This results in the following expressions for the post
collision state of the oscillating plate
\begin{align}
  t_{\rm shift}^\prime&=\frac{1}{\omega}\arctan \left( \frac{-\hat{n}
      \cdot \left[\Delta \vec{p}_{\rm Box} +\dot{\vec{r}}_{\rm
          Box}\right]}{2\, \pi\, \omega\, \hat{n} \cdot
      \left[\vec{r}_{\rm Box} -
        \vec{r}_{\rm Box}^{\,(0)}\right]} \right) - t \\
  \Delta^\prime &= \frac{\hat{n} \cdot \left(\vec{r}_{\rm Box} -
      \vec{r}_{\rm Box}^{\,(0)}\right)}{\cos\left(2\, \pi\,
      \omega\left[t + t_{\rm shift}^\prime\right]\right)}
\end{align}
where the primes denote post-collision values. Care must be taken at
this point in the calculation to ensure that the magnitude of $t$ and
$t_{\rm shift}$ do not affect the precision of the calculations. Care
must be taken also to retain the correct quadrant of the calculated
angle when using the $\arctan$ function.

A difficulty with the event-driven simulation method arises from its
inability to simulate events with finite durations. When the
oscillating wall is accelerating, a particle can repeatedly collide
with the plate until its relative velocity and separation are
numerically zero. Physically, the particle sticks to the wall and is
pushed until the plate enters the deceleration phase of its
oscillation, or interacts with another particle. To prevent this
unresolvable situation from occurring within the event-driven
simulation, the interactions between the oscillating wall and a
particle are turned elastic when
\begin{align}\label{eq:elasticlimit}
  \frac{\hat{n} \cdot \left(\vec{v}_i-\dot{\vec{r}}_{\rm
        Box}\right)}{\pi\, \omega\,\Delta} < 0.04
\end{align}
The pushing of the particle is then transformed into a sequence of
small hops which, as in the physical pushed case, do not
dissipate energy. As this expression is linear in the current
displacement $\Delta$, the long time behavior of the system is still
recovered ($\Delta\to0$ as $t\to\infty$). This elastic approximation
is small when the plate motion dominates the dynamics of the system
and the results appear to be unaffected if smaller values for
Eq.~\eqref{eq:elasticlimit} are used.
%%%%%%%%%%%%%%%%%%%%%%%%%%%%%%%%%%%%%%%%%%%%%%%%%%%%%%%%%%%%%%%%%%%%
\subsection{Parameters of the Simulation\label{sec:parameters}}
The simulations are initialized with all particles arranged in a
regular lattice (FCC), with initial velocities assigned from a
Gaussian and a total particle energy less than $<0.002\%$ of the
initial spring energy. The particles are packed in a loose layer on
the wall at the initial extrema of the oscillation. The particles in
the experiment are also typically arranged this way due to the
influence of gravity before the microgravity phase of the experiment.

The simulations require several inputs and these parameters are
reported in Table~\ref{tab:sysparam}.  All parameters, with the
exception of the box frequency $\omega$ and inelasticities
$\varepsilon_{\rm pp}$ and $\varepsilon_{\rm pw}$, are directly
obtained from the experimental setup described in
Sec.~\ref{sec:experiment}. The three remaining parameters must be
calculated from material parameters or obtained through experimental
results.

%%%%%%%%%%%%%%%%%%%%%%%%%%%%%%%%%%%%%%%%%%%%%%%%%%%%%%%%%%%%%%%%%%%%%%%%
%%%%%%%%%%%%%%%%%%%%%%%%%%%%%%%%%%%%%%%%%%%%%%%%%%%%%%%%%%%%%%%%%%%%%%%%
\begin{table}[h]
  \begin{center}
    \caption{\label{tab:sysparam} 
      Model parameters for the event-driven simulations.  }
    \begin{tabular*}{\columnwidth}{@{\extracolsep{\fill}}|c|c|c|c|c|c|c|c|}
      \hline\hline
      $\sigma$ (mm) & $m$ (g) & $N$ & $\Delta_0$ (mm)  & $\omega$ (s$^{-1}$) & $M$ (g) & $\varepsilon_{\rm pp}$ & $\varepsilon_{\rm pw}$ \\
      \hline
      10 & 4.04 & 37 & 107.5 & 1.23 & 434  & 0.75 & 0.76 \\
      \hline\hline
    \end{tabular*}
  \end{center}
\end{table}
% and~\ref{tab:sysdimensions}.

The frequency of the unloaded damper $\omega$ may be estimated using
the simple harmonic oscillator model. The spring constant of the
spring-blade may be calculated using the Euler-Bernoulli beam equation
\begin{align}
  k=\frac{E\,w\,h^3}{4\,l^3}=0.0254\,{\rm N}\,{\rm mm}^{-1}
\end{align}
where $E=2.06\times{}10^5$~N~mm${}^{-2}$ is the elastic modulus of the
spring steel, $w=30$~mm is the spring width, $h=1.5$~mm is the spring
%%%%%% length is -50mm for bottom clamp and -7.5mm for top clamp
thickness, and $l=590$~mm is the spring length. If the system behaves
as a simple harmonic oscillator and the mass of the spring is ignored
the frequency may be estimated using
\begin{align}
  \omega\approx \frac{1}{2\,\pi}\sqrt{\frac{k}{M}}\approx 1.217\,{\rm s}^{-1}
\end{align}
The frequency of the loaded granular damper ($\omega_{\rm system}$) is
lower than that of the empty damper ($\omega$) due to the added mass
and the interactions of the granulate. In the simple harmonic
oscillator model, the additional mass of the granulate alters the
frequency of the oscillations by
\begin{align}\label{eq:freqRelation}
  \omega_{\rm{}loaded}= \omega \sqrt{\frac{M}{M+N\,m}}
\end{align}
In the limit that the granulate is tightly packed in the granular
damper, the frequency of the system should limit to the simple
harmonic oscillator frequency
$\omega_{\rm{}system}\to\omega_{\rm{}loaded}$. In the limit of a large
box, the granulate will completely decouple from the oscillator and
$\omega_{\rm{}system}\to\omega$. Remarkably, the frequency of the
experimental oscillators, obtained through averaging the peak and
center point frequencies, is consistent for all box lengths at
approximately $\omega_{\rm{}system}\approx{}1.05\,{\rm{}s}^{-1}$ with
a standard deviation of $\pm0.01\,{\rm{}s}^{-1}$. If its assumed that
$\omega_{\rm{}system}\approx\omega_{\rm{}loaded}$ for small box
lengths, Eq.~\eqref{eq:freqRelation} estimates an unloaded frequency
of $\omega\approx1.22\pm0.1\,{\rm{}s}^{-1}$ for the experimental
system. This agreement with the beam equation is promising and
suggests that, although the granulate is periodically decoupled from
the oscillator, the deviation from Eq.~\eqref{eq:freqRelation} is
still small for the experimental box lengths studied here.
For the simulations, a slightly higher frequency of
$\omega\approx1.23\,\,{\rm{}s}^{-1}$ is used which is within the
standard deviation of the experimental values and yields an excellent
fit to the experimental data.

Finally, the coefficients of restitution $\varepsilon_{\rm pp}$ and
$\varepsilon_{\rm pw}$ describing the inelastic collisions between
particles and between a particle and the wall must be
determined. These model parameters are obtained by fitting simulation
results to the experimental data for the smallest box length ($L=40$
mm, Fig.~\ref{fig:comp1}).
As best fits the following results are obtained
\begin{align}\label{eq:3}
  \varepsilon_{\rm pp}&=0.75& \varepsilon_{\rm pw}&=0.76
\end{align}
The value for the particle-wall coefficient of restitution is in close
agreement with published results reported for a 9.35mm steel
ball-bearing impacting a clamped acrylic
plate~\cite{Sondergaard_1990}; however, the particle-particle
inelastic coefficient is significantly lower than expected.
Performing an automated drop test~\cite{MARINA_MICHEAL_PAPER} of the
granulate on to a silicon carbide plate yields an elasticity of
$\varepsilon\approx0.95$.  Due to the high rigidity of the base plate,
this value should be close to the experimental value for
particle-particle interactions. The fitted particle-particle
inelasticity $\varepsilon_{\rm pp}$ may be unexpectedly lower than the
drop test results due to missing dissipation mechanisms in the model
(e.g., granulate friction). Despite this, the agreement of the
simulation and experimental results (see Sec.~\ref{sec:validation})
shows that this is still an effective model for the system.

It should be noted that the optimization/fitting of the inelasticities
$\varepsilon_{\rm pp}$ and $\varepsilon_{\rm pw}$ is performed
exclusively for the box width of $L=40$ mm. For all other simulations
reported here, the optimal coefficients of restitution are used
without further fitting.  

%%%%%%%%%%%%%%%%%%%%%%%%%%%%%%%%%%%%%%%%%%%%%%%%%%%%%%%%%%%%%%%%%%%%
\subsection{Validation of the Numerical Method\label{sec:validation}}
\begin{figure}[ht]
  \begin{center}
    \includegraphics[clip,width=\columnwidth]{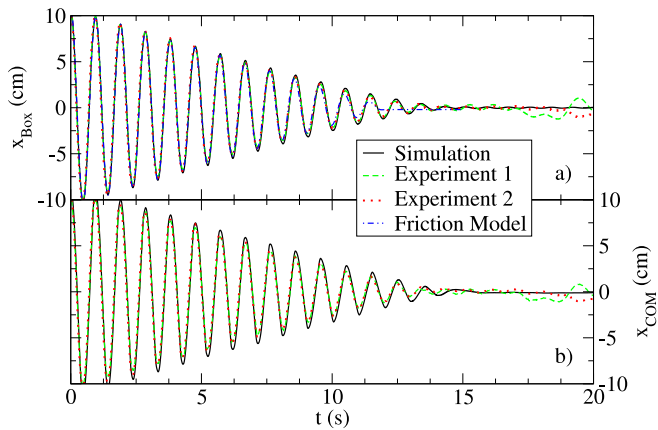}
    \caption{\label{fig:comp1} (Color online) A comparison of simulation results,
      experimental data and Eq.~\eqref{eq:frictDamper} for (a) the box
      position $x_{\rm Box}$ and (b) the granulate center of mass
      $x_{\rm COM}$ as a function of time for a box length of
      $L=40$~mm. The simulation data is fitted to the experimental
      data through the inelasticities $\varepsilon_{\rm pp}=0.75$ and
      $\varepsilon_{\rm pw}=0.76$.}
  \end{center}
\end{figure}

The simulation and experimental results are compared in this section
to validate the model. Figure~\ref{fig:comp1} presents the box
position $x_{\rm Box}$ and granulate center of mass $x_{\rm COM}$ as a
function of time for a box length of $L=40$~mm. Two experimental
measurements are reported and both are in close agreement with the
simulation results. The experimental and simulation results display a
high degree of repeatability and single realizations are
representative of the averaged values. This is due to the uniqueness
of the initial state, with the spring held in a deflected state and
the particles resting in a regular, repeatable layer on the outer wall
due to the influence of gravity before the microgravity
phase. However, the experimental results begin to fluctuate towards
the end of the microgravity phase due to disturbances in the flight.

The numerical result for the box position $x_{\rm Box}$ as a function
of time is in excellent quantitative agreement with the experimental
data. For the position of the center of mass, $x_{\rm COM}$ the
agreement is also very good albeit not as close as for
$x_{\rm{}Box}(t)$, with some over-estimations near the peaks of the
oscillations. The error could arise from the experimental method due
to the top-down view of the simulation and 2D image reconstruction
used. The area of the visible particles are identified and the
centroid location is taken to be the center of mass. Due to the end
walls and slight arcing motion of the box (see
Fig.~\ref{fig:experimentaldesign}) the reconstructed center of mass is
slightly biased towards the center of the box.

\begin{figure}[ht]
  \begin{center}
    \includegraphics[clip,width=\columnwidth]{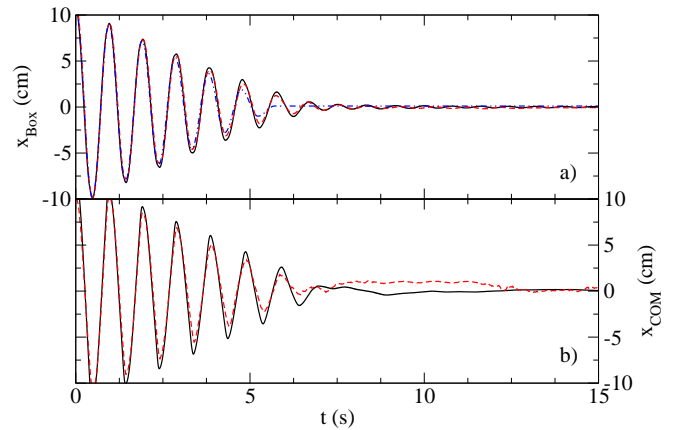}
    \caption{\label{fig:comp2} (Color online) A comparison of simulation results,
      experimental data and Eq.~\eqref{eq:frictDamper} for (a) the box
      position $x_{\rm Box}$ and (b) the granulate center of mass
      $x_{\rm COM}$ as a function of time for a box length of
      $L=40$~mm. Line types are described in Fig.~\ref{fig:comp1}.
      The simulation data is not fitted to the experiment and the
      parameters of Fig.~\ref{fig:comp1} are used.}
  \end{center}
\end{figure}

\begin{figure}[ht]
  \begin{center}
    \includegraphics[clip,width=\columnwidth]{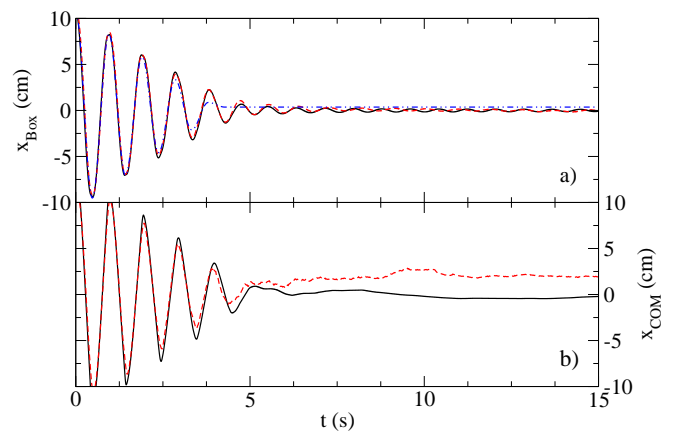}
    \caption{\label{fig:comp3} (Color online) The same comparison as
      Fig.~\ref{fig:comp2}, but for a box length of $L=85$~mm.}
  \end{center}
\end{figure}

\begin{figure}[ht]
  \begin{center}
    \includegraphics[clip,width=\columnwidth]{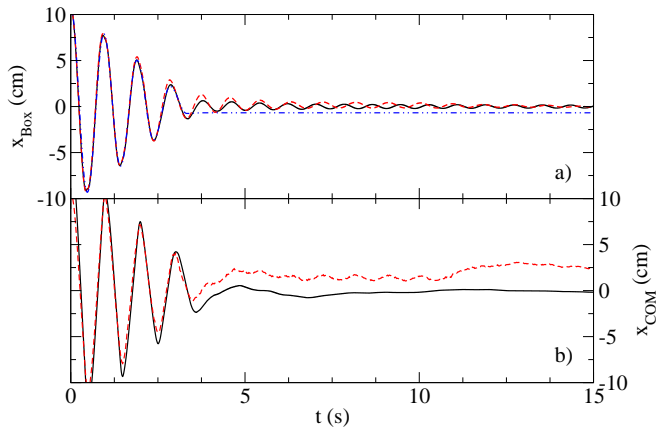}
    \caption{\label{fig:comp4} (Color online) The same comparison as
      Fig.~\ref{fig:comp2}, but for a box length of $L=104$~mm.}
  \end{center}
\end{figure}

The agreement between the simulation and experiment for the frequency
of the damped oscillator is excellent and confirms the accuracy of the
fundamental frequency $\omega$; however, the excellent agreement in
the amplitudes between experimental data and simulations for $L=40$~mm
is perhaps not too surprising since this experimental data set is used
to determine the coefficients of restitution, $\varepsilon_{\rm pp}$
and $\varepsilon_{\rm pw}$. The model parameters are now fixed and the
numerical result for several different box widths are compared with
the corresponding experimental data (see
Figs.~\ref{fig:comp2}-\ref{fig:comp4}).

In general, the simulation results are in excellent agreement with the
models predictions. This implies that the approximations of the model
(1~D oscillations, no air resistance, ideal spring) are small and have
little effect on the dynamics of the granular damper. Some of these
approximations may already be compensated for in the fitting of the
coefficients of restitution, but they appear to be well behaved with
the changes in box length. In the simulation, rotational degrees of
freedom are neglected by eliminating friction between the particles
and the particles and the container walls. In contrast to vibrated
granular dampers in gravity, where the energy dissipation due to
friction is of the same scale as energy dissipation by
impact~\cite{ChenEtAl:2001}, in microgravity friction seems to be less
important or easily characterized into the restitution coefficient
$\varepsilon_{\rm{}pp}$. Overall, the fitting of the inelasticities
appears to be effective at capturing the behavior of the system and no
further parameters or extensions of the simulation model are required.

The most striking feature of the curves in
Figs.~\ref{fig:comp1}-\ref{fig:comp4} is the linear decay of the peak
amplitude of the oscillation with time. A detailed discussion of this
property is postponed to Sec.~\ref{sec:macmodel} and optimal dampers
are discussed in the following section.
%%%%%%%%%%%%%%%%%%%%%%%%%%%%%%%%%%%%%%%%%%%%%%%%%%%%%%%%%%%%%%%%%%%%
%%%%%%%%%%%%%%%%%%%%%%%%%%%%%%%%%%%%%%%%%%%%%%%%%%%%%%%%%%%%%%%%%%%%
\section{Optimal Dampers\label{sec:optimisation}}
There is a significant dependence of the damping efficiency on the
container length, as is seen in
Figs.~\ref{fig:comp1}--\ref{fig:comp4}. The number of cycles before
the oscillations are sufficiently damped varies from 13 to 4 as the
box length is increased. By examining the energy transfer mechanisms
within the granular damper, an expression for optimizing the dampers
design may be found.

\begin{figure}[ht]
  \begin{center}
    \includegraphics[clip,width=\columnwidth]{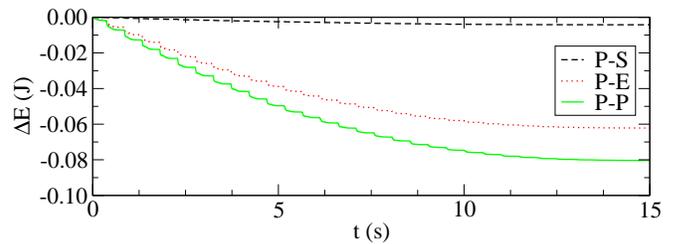}
\caption{\label{fig:EnergyLoss} (Color online)
  Simulation values for the cumulative energy loss through
  interactions with the side walls (P-S), end walls (P-E) and
  particle-particle interactions (P-P) for a box length of $L= 40$ mm.}
\end{center}
\end{figure}

Figure~\ref{fig:EnergyLoss} plots the cumulative energy lost through
the three classes of collisions in the simulation system. It should be
noted that Fig.~\ref{fig:EnergyLoss} is only valid for the fitted
inelasticities and will therefore differ from the true experimental
values. Nevertheless, the results should agree qualitatively and allow
some insight into the experimental system.  The sides of the box
appear to be unimportant in this design of a damper and may present an
opportunity for optimization by utilizing alternative shaker
geometries (e.g., an hourglass design).  Not only are the particle-end
wall collisions the sole mechanism for the transferal of oscillation
energy from the oscillator into the contained granulate, but
simulation results estimate that these collisions are also a
significant dissipation mechanism for the damper.  The end wall
interactions both transfer and dissipate the maximum energy when the
relative velocity of the oscillator end walls and granulate are
maximized. Therefore, maximizing this relative speed should optimize
the performance of the granular damper. In the following subsection,
an attempt is made to estimate the optimal box length using a simple
model for the dynamics.
%
%%%%%%%%%%%%%%%%%%%%%%%%%%%%%%%%%%%%%%%%%%%%%%%%%%%%%%%%%%%%%%%%%%%%
\subsection{Theoretical Predictions\label{sec:optimisation.theory}}
Attempting to optimize the system by modeling the granulate as a
single particle or some other simplified description is
difficult~\cite{Popplewell_Liao_1991} due to the lack of an analytical
solution to Eq.~\eqref{eq:f0}.  To estimate the optimal damping
conditions, only two plausible assumptions for the motion of the
granulate in the box are required: a) the granulate will be
``collected'' and form a packed layer on the approaching oscillating
wall during the initial inward stroke (when the oscillator accelerates
towards the center point), and b) the center of mass velocity of the
granulate at the end of the inward stroke is, on average, the maximum
oscillator velocity.  The time from the center of the stroke at which
the granulate would hit the peak displacement of the oscillator is
then given by
\begin{align}\label{eq:tgpeak}
  t_{\rm g,peak} = \frac{L + \Delta - \sigma_{\rm layer}}{2
    \,\pi\,\omega_{\rm loaded}\,\Delta}
\end{align}
where $\sigma_{\rm layer}=20$~mm is the thickness of the layer of
granulate when it is packed on the surface of the oscillating wall. It
should be noted that Eq.~\eqref{eq:tgpeak} decreases in time, as
$\Delta$ decreases on average due to interactions with the granulate.
If for any integer $n$ the peak collision time lies in the range
$n<\omega\,t_{\rm{}g,peak}<n+\nicefrac{1}{4}$, the granulate will hit
the oscillating wall on the outward phase of its stroke. All
experimental box with the exception of the largest system ($L=104$~mm)
are within this regime. It is expected that improved damping occurs if
the granulate hits on the inward stroke as the relative velocity is
maximized.

The granulate travels the length of the box in
\begin{align}\label{eq:optboxeq}
  t_{\rm g,Box} = \frac{L - \sigma_{\rm
      layer}}{2\,\pi\,\omega_{\rm loaded}\,\Delta}
\end{align}
If $n+\nicefrac{1}{4}>\omega\,t_{\rm{}g,peak}>n+1$ and
$n<\omega\,t_{\rm{}g,Box}<n+\nicefrac{1}{2}$, the granulate will
collide on the inward phase of the stroke. The largest system, where
$\omega\,t_{\rm{}g,Box}\approx0.15$, collides after the turning point
of the oscillator; however, the dissipation is maximized when
$\omega\,t_{\rm{}g,Box}\approx\nicefrac{1}{2}$.  At this point, the
relative velocity between the granulate and oscillating box is also
maximized.  For $\omega\,t_{\rm{}g,Box}>\nicefrac{1}{2}$ the plate is
either decelerating or multiple cycles of the oscillation occur
without the granulate colliding.

The damping of the oscillator from the initial state can be optimized,
independently of the inelastic coefficients, by altering either $L$,
$\omega$, or $\Delta$ such that
$\omega\,t_{\rm{}g,Box}\approx\nicefrac{1}{2}$. Efficiency will be
lost and recovered as $\Delta$ changes over time, but if the granulate
is relatively inelastic this will occur after most of the energy is
dissipated or transferred in the first cycle.

Setting $\omega\,t_{\rm{}g,Box}=\nicefrac{1}{2}$ in
Eq.~\eqref{eq:optboxeq} and using Eq.~\eqref{eq:freqRelation}, the
optimal box length $L_{\rm opt}$ may be estimated for a given initial
amplitude $\Delta_0$ using
\begin{align}\label{eq:optimalBoxLength}
  L_{\rm opt}&= \pi\,\Delta_0\sqrt{\frac{M}{M+N\,m}} +\sigma_{\rm layer}
\end{align}
This expression is remarkable in that it is independent of the
oscillation frequency. This may be understood from dimensional
analysis as, due to the negligible initial kinetic energy, the model
has only one time scale. As such, the solutions to the model must
scale trivially in the frequency of the oscillations. In the following
subsection, the results of Eq.~\eqref{eq:optimalBoxLength} and its
assumptions are checked against simulation results.
\subsection{Numerical Test\label{sec:optimisation.MD}}
\begin{figure}[ht]
  \begin{center}
    \includegraphics[clip,width=\columnwidth]{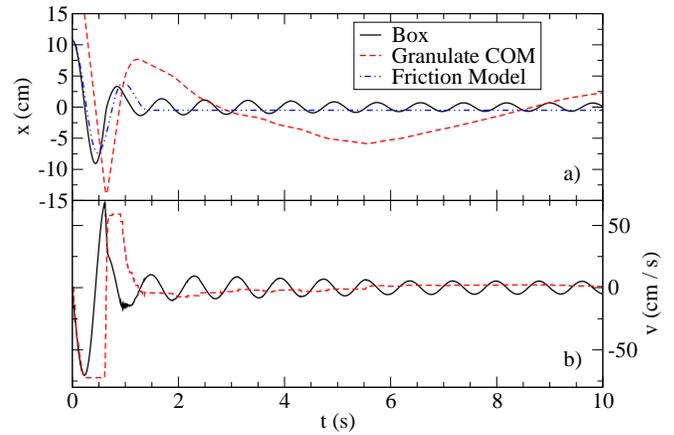}
    \caption{\label{fig:optimised} (Color online)
      A comparison of simulation results, experimental data and
      Eq.~\eqref{eq:frictDamper} for the box and granulate (a)
      position, and (b) velocity as a function of time for the optimal
      box length of $L=311$~mm, as predicted by
      Eq.~\eqref{eq:optimalBoxLength}.}
  \end{center}
\end{figure}

The validity of the basic assumptions made in
Sec.~\ref{sec:optimisation.theory} and the result,
Eq.~\eqref{eq:optimalBoxLength}, are now tested using the results of
the DEM simulations.  Using Eq.~\eqref{eq:optimalBoxLength} to predict
the optimal box length for the damping of the experimental system
yields a value of $L_{\rm{}opt}=311$~mm.  The results of a simulation
at this box length are presented in Fig.~\ref{fig:optimised}.  A
square step in the granulate center of mass velocity is visible at the
peak of the box velocity as the granulate decouples from the
oscillator. The assumption of an equal box and granulate velocity at
the midpoint of the stroke (at peak velocity) appears to hold. Visual
inspection~\cite{sim_video_data} confirms the granulate is collected
in a layer on the approaching oscillating wall. The re-collision of
the granulate also appears to occur close to the peak of the box
velocity, maximizing the relative velocity, energy dissipation, and
energy transfer in this first collision. The largest oscillations are
effectively damped within one second; however, the oscillator is now
susceptible to smaller amplitude oscillations which appear to decay
very slowly. The optimal approach would be to couple two or more
dampers to damp a wider range of amplitudes within short
timescales. This idea has already been pursued for impact dampers
(e.g., Ref.~\cite{Masri_1969}) which are related to granular dampers
except that in the container or cavity there is only a single
particle.

\begin{figure}[ht]
  \begin{center}
    \includegraphics[clip,width=\columnwidth]{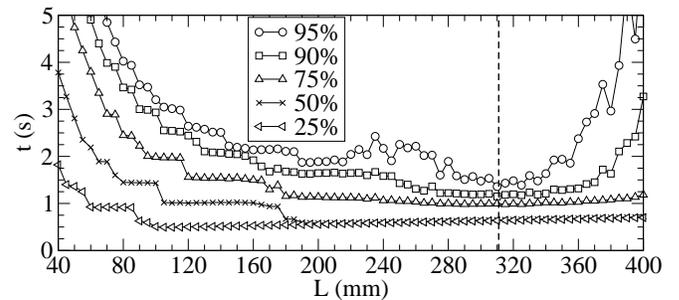}
    \caption{\label{fig:optimalLSweep} 
      Simulation results for the time $t$ to dissipate a percentage of
      the initial energy, versus the box length $L$. The vertical
      dashed line indicates the optimal box length as predicted by
      Eq.~\eqref{eq:optimalBoxLength}.}
  \end{center}
\end{figure}

To test the predictions of Eq.~\eqref{eq:optimalBoxLength} for the
optimal damping length $L$, a suitable metric must be defined to
compare various box lengths. Figure~\ref{fig:optimalLSweep} compares
the time an oscillator takes to dissipate a certain fraction of the
initial energy as a function of the box length. Despite the continuing
low-amplitude oscillations of the damper at $L=311$~mm (see
Fig.~\ref{fig:optimised}), the damper effectively eliminates 95\% of
the initial energy in well under two
oscillations. Equation~\eqref{eq:optimalBoxLength} appears to yield an
excellent estimate for the global optimal box length, avoiding both
the highly inefficient zones towards the edges of the graph. Previous
work on forced granular dampers (e.g., see Fig.~7 in
Ref.~\cite{Saeki:2002}) also yields performance curves with the same
general U-shape as Fig.~\ref{fig:optimalLSweep}.
An alternative metric for comparing the efficiency of granular dampers
is now defined through a phenomenological model for the damping
behavior.
\section{Phenomenological Model of Granular Dampers\label{sec:macmodel}}
\begin{figure}[ht]
  \begin{center}
    \includegraphics[clip,width=\columnwidth]{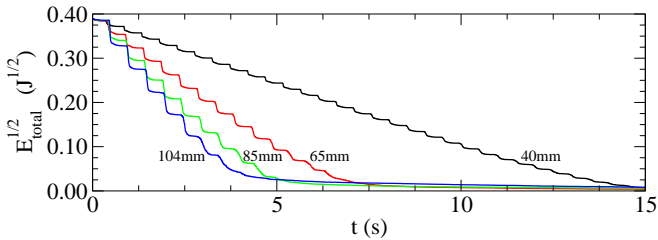}\vspace*{-0.3cm}
\caption{\label{fig:energies} (Color online)
  Total energies of the oscillators as obtained from numerical
  simulations.}
\end{center}
\end{figure}

Figures~\ref{fig:comp1}--\ref{fig:comp4} reveal a linear decay of the
amplitude of the oscillation with time, and thus the energy of the
system decays quadratically in time. This is highlighted in
Fig.~\ref{fig:energies}, where the time dependence of the square root
of the total energy of the damper is plotted. This result is
surprising considering the approximations of the previous section: the
oscillator appears to have a constant frequency for a given box
length, and the oscillator collects the granulate on a wall and then
collides the granulate in each half period. For the amplitude decay to
be linear, the energy dissipated in each of these ``collisions'' of
the granulate must then be proportional to the amplitude
$\Delta$. However, if an inelastic particle is given a velocity
proportional to the maximum plate velocity ($2\,\pi\,\omega\,\Delta$),
it will dissipate energy proportional to the square of the plate
amplitude ($\Delta^2$) for a given number of collisions.

The result is also surprising as the more common viscous dampers yield
an exponential decay of the amplitude; however, the only simple
damped-oscillator which displays a linear decay in the amplitude is
one damped by a constant magnitude force~\cite{Hudson_Finfgeld_1971}
(e.g., a friction-damped oscillator). The equation of motion for such
an oscillator is
\begin{align}\label{eq:frictDamper}
  M_{\rm tot}\,\ddot{x} = -k\,x-\mu \,M_{\rm{}tot} \sign\left(\dot{x}\right)
\end{align}
where $x$ is the oscillator position, $M_{\rm{}tot}=M+N\,m$ is the total
oscillating mass and $\mu\,M_{\rm{}tot}$ is the magnitude of the constant
frictional force. No simple analytical solution exists to this
equation although piecewise solutions may be
found~\cite{Hudson_Finfgeld_1971}. This model does not appear to be
appropriate for the granular damper due to the intermittent nature of
the damping force. For example, the steps in the damper energy
(Fig.~\ref{fig:energies}) arise from the decoupling of the damping
granulate and oscillator at the midpoint of the stroke (see
Fig.~\ref{fig:optimised}) and during these steps the oscillator
experiences no damping force. However, this model may still be useful
in characterizing and comparing the damping efficiencies of granular
dampers in microgravity through the effective frictional force
$\mu\,M_{\rm{}tot}$.

The effective frictional force of a experimental damper may be
estimated through the decay of the peak amplitude, as given by
\begin{align}
  %\left|x_n\right| &= \Delta_0 - \frac{(2\,n + 1 - (-1)^{2\,n})\mu}{k}\\
  \mu\,M_{\rm tot} &= \frac{k\left(\Delta_0 -\left|x_n\right|
    \right)}{2\,n + 1 - (-1)^{2\,n}}\label{eq:frictioncoeff}
\end{align}
where $n$ is the index of the amplitude peak and $\left|x_n\right|$ is
the absolute oscillator displacement for the $n$th peak. The peak
number $n$, used to calculate the effective friction coefficient,
should be odd to only sample the amplitude during the ``collection''
phase of the oscillation and should be as small as possible for
correct measurement of rapid dampers. The earliest value of $n$ which
satisfies these requirements is $n=3$, at a time of $t=1.5/\omega_{\rm
  system}$. Equation~\eqref{eq:frictioncoeff} is used to extract an
effective friction force for each experimental system and the
corresponding solutions to Eq.~\eqref{eq:frictDamper} are plotted in
Figs.~\ref{fig:comp1}--\ref{fig:comp4}. For the non-optimal dampers
the model fits the data well. Deviations begin to appear towards the
end of the oscillations as the ``collect and collide'' motion begins
to break down and the granulate spreads uniformly over the box. This
lends weight to the argument that the ``collect and collide'' motion
of the granulate is responsible for the apparent friction-damped
behavior. For the optimal damper (see Fig.~\ref{fig:optimised}), the
model does not fit as well. The discrepancy arises from the
oscillators frequency for this box length being significantly
different from the predictions of Eq.~\ref{eq:freqRelation}. Simple
spring models will no longer work and appear to be constrained to the
range ($t_{\rm{}g,peak}\lesssim\nicefrac{1}{4}$). A better fit may be
obtained by fitting the spring coefficient $k$, however this is
unsatisfactory as the results of the model cannot be compared between
systems. Another deficiency of the friction model is that it predicts
that the damper will come to a complete halt after a finite time. It
fails to capture the persistent small amplitude oscillations (see
Fig.~\ref{fig:optimised}). The friction deceleration, $\mu$, still
appears to be a useful value for comparing the damping efficiency of
sub-optimal granular dampers.

The quadratic decay of energy with time in a granular system attached
to a linear spring has been reported
before~\cite{FriendKinra:2000,Marhadi:2003,MarhadiKinra:2005,BaiKeerWangSnurr:2009}.
% FriendKinra:2000 Exp Marhadi:2003,MarhadiKinra:2005 Exp cantilever
% vertical oscillation -- linear decay check this:
% BaiKeerWangSnurr:2009
Surprisingly, the same behavior is found also for rather different
dampers such as thrust-based damping~\cite{Shah_etal_2009} and impact
dampers~\cite{MansourFilho:1974,YangEtAl:2002,Cheng:2003,DuWangZhuEtAl:2008}.
% impact damper mit small amount of fine metal powder
% \cite{DuWangZhuEtAl:2008}
However, this is not a general rule and other published results exist
(e.g., Ref.~\cite{MaoWangXuEtAl:2004}) where a non-linear decay of the
amplitude of the oscillation (possibly exponential) is found.
This work clarifies that this apparent frictional behavior may also
arise solely from the collisional granular dynamics and does not
necessarily arise from friction forces within the experimental
setup. This is evident as the simple model used in the simulations
reproduces the linear decay of the amplitude.

\section{Conclusions \label{sec:conclusions}}
In this paper, a method for performing controlled experiments on
granular damped oscillators in microgravity is outlined. High-speed
video capture and image-processing techniques are used to reconstruct
the motion of the oscillator to obtain accurate experimental results.
A simple hard sphere model and event-driven dynamics are also used to
generate quantitative results that compare well against the
experimental values. From the excellent agreement of the simulation
and experimental frequency, it appears that the damper frequency
responds like a simple harmonic oscillator to changes in load
(Eq.~\eqref{eq:freqRelation}) for short box lengths. This is
remarkable given the periodic decoupling of the granulate from the
spring and box. The simulation model scales trivially with the
frequency of the oscillations as, apart from the negligible initial
energy, the model has only one time scale. Further research is
required on experimental systems to determine the frequency dependence
of granular dampers and generalize the current model to these systems.

The straightforward design of these granular dampers yields a
remarkably simple expression for the optimal damping configuration
of the form of Eq.~\eqref{eq:optimalBoxLength}. Simulations at the
predicted optimal box length damp large amplitude oscillations
remarkably well (see Fig.~\ref{fig:optimised}) but are susceptible to
smaller amplitude disturbances. The final expression for the optimal
box length is independent of the oscillation frequency, which may be
understood through dimensional analysis of the model.

Unlike conventional viscous-damped systems, the granular damped system
studied here displays a linear decay in the amplitude. This behavior
is not intuitive and is a feature typical of friction-damped
oscillators. The simulation results and their excellent agreement with
experimental results strongly suggest that this effect arises solely
from the granular dynamics. The linear decay is a useful property as
it implies that a granular damper can completely damp oscillations
within a finite time; however, this is not the case as, at low
oscillation energies a transition occurs and the damping force is
significantly reduced. Further research is required in designing
dampers with a wider amplitude response by coupling multiple dampers
with different lengths. The internal geometries may also be optimized
to eliminate the decoupling of the granulate in the midpoint of the
stroke to create more effective dampers.

\section*{Acknowledgments}
The authors would like to acknowledge the German Science Foundation
(DFG) for funding via the grant SP608 and DLR for funding the
parabolic flight campaign. Thanks also go to the mechanical workshop
NW2 at the University of Bayreuth. Finally, the authors would like to
acknowledge the additional funding of the DFG through the Cluster of
Excellence Engineering of Advanced Materials in Erlangen.

\end{document}